\documentclass{article}

\begin{document}

\title{A different approach to anisotropic spherical collapse with shear and heat
radiation}
\author{B.V.Ivanov \\
Institute for Nuclear Research and Nuclear Energy, \\
Bulgarian Academy of Science, \\
Tzarigradsko Shausse 72, Sofia 1784, Bulgaria}
\maketitle

\begin{abstract}
In order to study the type of collapse, mentioned in the title, we introduce
a physically meaningful object, called the horizon function. It directly
enters the expressions for many of the stellar characteristics. The main
junction equation, which governs the collapse, transforms into a Riccati
equation with simple coefficients for the horizon function. We integrate
this equation in the geodesic case. The same is done in the general case
when one or another of the coefficients vanish. It is shown how to build
classes of star models in this formulation of the problem and simple
solutions are given.
\end{abstract}

\section{Introduction}

Gravitational collapse is one of the main topics in relativistic
astrophysics. The first example was given by Oppenheimer and Snyder \cite
{one} who studied the collapse of a dust cloud. It has energy density but no
pressures. Then followed studies of the collapse of perfect or anisotropic
fluid spheres \cite{two}. Such models have clearly defined boundary, where
the interior solution matches the exterior Schwarzschild solution. The main
junction condition is the vanishing of the radial pressure at the surface of
the star. The process of collapse, however, is highly dissipative, required
to account for the enormous binding energy of the resulting object \cite
{three}. Thus a more realistic scenario is the collapse with heat flow \cite
{four} or pure radiation \cite{five}. Spherical collapse is described in the
general case by a diagonal metric with three independent components, $g_{tt}$%
, $g_{rr}$ and $g_{\theta \theta }$. The exterior solution is the Vaidya
shining star \cite{six}. In the diffusion approximation (heat flow) the main
junction condition changes. At the boundary surface the radial pressure
should equal the heat flux. In the streaming approximation (pure radiation)
it remains the same - vanishing of the radial pressure. The main junction
condition is a non-linear differential equation in partial derivatives
(along radius and time) for the three metric components and follows from the
matching of the second fundamental forms. It is in fact a constraint that
reduces the number of independent metric components to two. It has the same
form both in the heat flow case and in the null radiation case. The other
junction conditions give expressions for different characteristics of the
exterior solution in terms of the interior one.

For simplicity, shearless fluid is discussed quite often, because then the
differential equation involves only the two components $g_{tt}$ and $g_{rr}$%
. When the fluid is perfect they have to satisfy also another differential
equation, called the isotropy condition. Even in this case the amount of
interior solutions is enormous \cite{seven}. Global solutions for perfect
fluids were given soon after the new junction equation was derived \cite
{eight}, \cite{nine}. Separation of variables allows to integrate the
equation and see explicitly the evolution of the star from a static model to
a black hole. A number of such solutions were found later, e.g. \cite{ten}.
It has been noticed that there is an additional, simple solution, linear in
time for $g_{rr}$, which leads to eternal collapse and a horizon never
appears. This happens both for perfect \cite{five}, \cite{seven}, \cite
{eleven} and imperfect fluids \cite{twelve}. An example of this no-horizon
phenomenon is given for the geodesic case ($g_{tt}=1$) without shear, too 
\cite{thirteen}. Another approach to the junction equation uses its Lie
point symmetries and provides different classes of solutions. This procedure
was applied to perfect fluid models \cite{fourteen} and to models with
anisotropic pressures \cite{fifteen}.

When shear is present, the general three component metric should be used,
which further complicates the differential equation. The geodesic case,
however, is simpler, involving again two components, $g_{rr}$ and $g_{\theta
\theta }$. The first exact geodesic solution with radiation was obtained by 
\cite{sixteen}. After that it was noticed \cite{seventeen} that the junction
condition is a Riccati equation for $g_{rr}$. It is of first order and is
not integrable in general. When some of its coefficients are set to zero, it
reduces to integrable equations and simple expressions for $g_{rr}$ may be
found. In this approach one finds first the metric, which is fixed, because
comoving coordinates are used, and then obtains the physical characteristics
of the model from the Einstein equations. Two simple regular solutions in
separated variables were derived. The previous solution is regained when
certain parameters are set to zero. Later, more general exact solutions,
depending on arbitrary functions of the coordinate radius, were given \cite
{eighteen}. They encompass the previous solutions. In a recent development
the realm of analytic solutions was further expanded by studying the Lie
point symmetries of the junction condition \cite{nineteen}. This results in
generalized traveling waves and self-similar solutions. Finally, all
solutions of the equation were found with the help of two generating
functions, based on the approach discussed in the present paper \cite{twenty}%
. All previous solutions were obtained by choosing the generating functions
in a proper way.

The general case with shear has been attacked both numerically and
analytically. Chan and co-workers used the initial method of separated
variables. In the presence of shear the time evolution equation is not
integrable and has to be solved numerically \cite{twone}, \cite{twtwo}, \cite
{twthree}, \cite{twfour}, \cite{twfive}. In some cases the initial static
model evolves into a black hole, but in others the star burns out
completely, radiating its mass away, so that no horizon ever forms and the
final result is flat spacetime.

The analytic approach is based on the fact that $g_{rr}$ satisfies a Riccati
equation, like in the geodesic case. Its reduction to a linear, Bernoulli,
or simpler Riccati equation is exploited to find solutions in elementary
functions \cite{twsix}, \cite{twseven}, \cite{tweight}.

Another class of shearing, radiative, anisotropic fluid models are the
so-called Euclidean stars \cite{twnine}. Their thermal behaviour was studied 
\cite{thirty}. A class of Euclidean stars with no horizon was found \cite
{thone}. Finally, the Lie symmetry approach has been applied to the general
case with shear and classes of new solutions were produced. They were called
generalized Euclidean stars since some of the usual Euclidean stars are
obtained as a subcase \cite{thtwo}.

L. Herrera and co-authors have studied different aspects of relativistic
stars with shear, like expansionfree collapse \cite{ththree}, the evolution
equation of the shear and the extension of the mass formalism to the
radiative case \cite{thfour}, the cause of energy-density inhomogeneity \cite
{thfive} and the collapse in the post-quasistatic approximation \cite{thsix}.

The main idea of the approach used in the present paper is to transform the
junction equation into an equation for a physically meaningful object, which
we call the horizon function. It is directly related to the redshift and the
formation of a horizon, which means the appearance of a black hole as the
end product of collapse. It enters the expression for the mass of the star,
the heat flow and the luminosity at infinity.

In Sect. 2 we present the Einstein equations, which in the anisotropic case
are expressions for the energy density, the radial and the tangential
pressure and the heat flow. The definitions of the shear, the expansion, the
horizon function and the redshift are given. The relation between the mass
of the star and the horizon function is clarified. The main results of the
matching to the exterior Vaidya solution are shown. The most important of
them is a differential equation involving the metric components. We show
that it is the same in the diffusion and in the streaming approximations.
The luminosities at the star's surface and at infinity are written, as well
as the surface temperature. We also use another version of the Einstein
equations, known as the mass formalism, to give additional expressions for
the heat flow and the energy density, which are connected to the mass. In
Sect. 3 the junction equation is written as a Riccati equation for the
horizon function with simple coefficients and the process of building a star
model solution is clarified. In Sect. 4 all geodesic solutions are derived
from a single generating function. In Sect. 5 the general Riccati equation
is reduced to a special one, without a linear term, which is integrable. The
simplest solution is demonstrated. In Sect. 6 the general Riccati equation
is reduced to a linear one, which is integrated too. A solution, for a
linear in time star radius is studied. Sect. 7 represents a discussion.

\section{Stellar characteristics}

The collapse of an anisotropic fluid sphere with shear is described by the
following metric 
\begin{equation}
ds^2=-A^2dt^2+B^2dr^2+R^2\left( d\theta ^2+\sin ^2\theta d\varphi ^2\right) ,
\label{one}
\end{equation}
where $A$, $B$ and $R$ are independent functions of time $t$ and the radius $%
r$ only. The spherical coordinates are numbered as $x^0=t$, $x^1=r$, $%
x^2=\theta $ and $x^3=\varphi $. The energy-momentum tensor, describing
dissipation through heat flow and null fluid, reads 
\begin{equation}
T_{ik}=\left( \mu +p_t\right) u_iu_k+p_tg_{ik}+\left( p_r-p_t\right) \chi
_i\chi _k+q_iu_k+u_iq_k+\varepsilon l_il_k.  \label{two}
\end{equation}
Here $\mu $ is the energy density, $p_r$ is the radial pressure, $p_t$ is
the tangential pressure, $u^i$ is the four-velocity of the fluid, $\chi ^i$
is a unit spacelike vector along the radial direction, $q^i$ is the heat
flow vector, also in the radial direction, $\varepsilon $ is the energy
density of the null fluid and the vector $l^i$ is null. In comoving
coordinates we have 
\begin{equation}
u^i=\delta _0^i,\quad \chi ^i=B^{-1}\delta _1^i,\quad q^i=q\chi ^i,\quad
l^i=u^i+\chi ^i.  \label{three}
\end{equation}
The Einstein field equations read 
\begin{equation}
\mu +\varepsilon =\frac 1{A^2}\left( \frac{2\dot B}B+\frac{\dot R}R\right) 
\frac{\dot R}R-\frac 1{B^2}\left( \frac{2R^{\prime \prime }}R+\frac{%
R^{\prime 2}}{R^2}-\frac{2B^{\prime }R^{\prime }}{BR}-\frac{B^2}{R^2}\right)
,  \label{four}
\end{equation}
\begin{equation}
p_r+\varepsilon =-\frac 1{A^2}\left[ \frac{2\ddot R}R-\left( \frac{2\dot A}A-%
\frac{\dot R}R\right) \frac{\dot R}R\right] +\frac 1{B^2}\left( \frac{%
2A^{\prime }}A+\frac{R^{\prime }}R\right) \frac{R^{\prime }}R-\frac 1{R^2},
\label{five}
\end{equation}

\begin{eqnarray}
p_t &=&-\frac 1{A^2}\left[ \frac{\ddot B}B+\frac{\ddot R}R-\frac{\dot A}%
A\left( \frac{\dot B}B+\frac{\dot R}R\right) +\frac{\dot B\dot R}{BR}\right]
+  \nonumber  \label{six} \\
&&\ +\frac 1{B^2}\left[ \frac{A^{\prime \prime }}A+\frac{R^{\prime \prime }}%
R-\frac{A^{\prime }B^{\prime }}{AB}+\left( \frac{A^{\prime }}A-\frac{%
B^{\prime }}B\right) \frac{R^{\prime }}R\right] ,  \label{six}
\end{eqnarray}
\begin{equation}
qB+\varepsilon =-\frac 2{AB}\left( \frac{\dot BR^{\prime }}{BR}-\frac{\dot
R^{\prime }}R+\frac{\dot RA^{\prime }}{RA}\right) ,  \label{seven}
\end{equation}
Here the dot means a time derivative, while the prime stands for a radial
derivative.

For the line element (1) the shear $\sigma $, the four-acceleration $a_1$
and the expansion scalar $\Theta $ are given by 
\begin{equation}
\sigma =\frac 13\left( \frac{\dot R}R-\frac{\dot B}B\right) ,\quad
\label{eight}
\end{equation}
\begin{equation}
a_1=\frac{A^{\prime }}A,\quad \Theta =\frac{2\dot R}R+\frac{\dot B}B.
\label{nine}
\end{equation}
Next, we introduce the important object $H$, which we call ''the horizon
function'' for reasons to become clear later: 
\begin{equation}
H=\frac{R^{\prime }}B+\frac{\dot R}A.  \label{ten}
\end{equation}
The mass $m,$ entrapped within radius $r$ is given by the expression \cite
{thseven} 
\begin{equation}
m=\frac R2\left[ 1+\left( \frac{\dot R}A\right) ^2-\left( \frac{R^{\prime }}%
B\right) ^2\right] .  \label{eleven}
\end{equation}
On the stellar surface $\Sigma $ it becomes the mass of the star. The
compactness parameter reads $u=m/R$. Eq (11) can be rewritten using $H$%
\begin{equation}
\frac{2m}R=1-H^2+\frac{2\dot R}AH.  \label{twelve}
\end{equation}

The exterior spacetime is given by the Vaidya shining star solution 
\begin{equation}
ds^2=-\left[ 1-\frac{2M\left( v\right) }\rho \right] dv^2-2dvd\rho +\rho
^2\left( d\theta ^2+\sin ^2\theta d\varphi ^2\right) ,  \label{thirt}
\end{equation}
where $M\left( v\right) $ is the mass of the star measured at time $v$ by an
observer at infinity, while $\rho $ is the outer radius. Both solutions
should be joined smoothly at $\Sigma $, which leads to the following
junction conditions \cite{theight}, \cite{thnine}: 
\begin{equation}
R_\Sigma =\rho _\Sigma \left( v\right) ,  \label{fourt}
\end{equation}
\begin{equation}
m_\Sigma =M_\Sigma ,  \label{fift}
\end{equation}
\begin{equation}
\left( p_r\right) _\Sigma =\left( qB\right) _\Sigma .  \label{sixt}
\end{equation}
Eq (16) should be satisfied by $A,$ $B$ and $R$, while the other equations
are definitions of different stellar characteristics. When there is only
null radiation, $q$ vanishes and the radial pressure should vanish at the
surface. This condition was used in many works with null fluid radiation 
\cite{five}. However, when $\varepsilon \neq 0$ Eqs.(5,7) show that in terms
of the metric Eq (16) is restored. Thus it is universal for heat and null
radiation. Therefore, in the following we set $\varepsilon =0$ and study
only the heat radiation case. Now when $q$ vanishes we get collapse without
radiation and the exterior solution is the Schwarzschild vacuum solution.

Some important stellar characteristics are defined on the surface of the
star. These are the redshift $z_\Sigma $%
\begin{equation}
z_\Sigma =\frac 1{H_\Sigma }-1,  \label{sevent}
\end{equation}
the surface luminosity $\Lambda _\Sigma $ and the luminosity at infinity $%
\Lambda _\infty $%
\begin{equation}
\Lambda _\Sigma =\left( \frac 12qBR^2\right) _\Sigma ,  \label{eighteen}
\end{equation}
\begin{equation}
\Lambda _\infty =H_\Sigma ^2\Lambda _\Sigma .  \label{ninet}
\end{equation}
The temperature at the surface is given by 
\begin{equation}
T_\Sigma ^4=\frac{\left( qB\right) _\Sigma }{8\pi \delta },  \label{twenty}
\end{equation}
where $\delta $ is some constant.

Additional formulas are obtained from the mass formalism \cite{forty}. In
its original form it gives expressions for the derivatives of the mass
function in the non-radiative case. It has been expanded to the dissipative
case \cite{thfour} 
\begin{equation}
\frac{\dot m}A=-\frac 12\left( p_r\frac{\dot R}A+qB\frac{R^{\prime }}%
B\right) R^2,  \label{twone}
\end{equation}

\begin{equation}
\frac{m^{\prime }}{R^{\prime }}=\frac 12\left( \mu +qB\frac{B\dot R}{%
AR^{\prime }}\right) R^2.  \label{twtwo}
\end{equation}
On the star's surface Eq (21) becomes, using the junction condition (16), 
\begin{equation}
\left( \dot m=-\frac 12qBHAR^2\right) _\Sigma .  \label{twthree}
\end{equation}
Collapse takes place with heat radiated towards outside, hence, $q>0$. Then $%
m$ decreases since energy is lost. Eq (22) gives another formula for the
energy density throughout the star 
\begin{equation}
\mu =\frac{2m^{\prime }}{R^2R^{\prime }}-\frac{qB^2\dot RR^2}{AR^{\prime }}.
\label{twfour}
\end{equation}

It is seen that the star properties have simpler expressions when written in
terms of $H$. The redshift is positive during collapse. Then Eq (17) shows
that $0\leq H_\Sigma \leq 1$. When $H_\Sigma =0$ we obtain from Eq (12) and
the junction conditions 
\begin{equation}
\left( 1-\frac{2m}R\right) _\Sigma =\left( 1-\frac{2M\left( \nu \right) }%
\rho \right) _\Sigma =0.  \label{twfive}
\end{equation}
This signals the appearance of a horizon and a black hole within it, which
is the typical end of gravitational collapse. This explains why we call $H$
the horizon function. The redshift becomes infinite, while the luminosity at
infinity drops to zero. The point in time when collapse starts is taken as $%
t_i$ (usually it is $-\infty $). There $H_\Sigma $ should have some positive
value $H_{\Sigma i},$ less or equal to $1$. Thus during the collapse to a
black hole the horizon function decreases to zero and $\dot H_\Sigma \leq 0$.

\section{Simplification of the junction equation}

With the help of Eqs (5,7) the main junction Eq (16) becomes on the star
surface 
\begin{eqnarray}
-\frac 2{AB}\left( \frac{\dot BR^{\prime }}{Br}-\frac{\dot R^{\prime }}R+%
\frac{\dot RA^{\prime }}{RA}\right) &=&-\frac 1{A^2}\left[ \frac{2\ddot R}%
R-\left( \frac{2\dot A}A-\frac{\dot R}R\right) \frac{\dot R}R\right] + 
\nonumber  \label{twsix} \\
&&+\frac 1{B^2}\left( \frac{2A^{\prime }}A+\frac{R^{\prime }}R\right) \frac{%
R^{\prime }}R-\frac 1{R^2}.  \label{twsix}
\end{eqnarray}
It is not hard to see that this is a Riccati equation for $B$. Regrouping
the terms we get 
\begin{equation}
\dot B=\left( \frac{2\ddot R}R+\frac{\dot R^2}{R^2}-\frac{2\dot A\dot R}{AR}+%
\frac{A^2}{R^2}\right) \frac R{2AR^{\prime }}B^2+\left( \frac{\dot R^{\prime
}}{R^{\prime }}-\frac{A^{\prime }\dot R}{AR^{\prime }}\right) B-\frac
A2\left( \frac{R^{\prime }}R+\frac{2A^{\prime }}A\right) .  \label{twseven}
\end{equation}
This is exactly Eq(11) from Ref. \cite{twsix}. There is no general solution
of the Riccati equation, although it is integrable in many concrete cases.
Furthermore, when the coefficient in front of $B^2$ vanishes it becomes a
linear equation, which is solvable for any given $A$ and $R$. When the free
term vanishes it becomes a Bernoulli equation, which is also solvable.
Finally, when the coefficient before the linear term vanishes, we get
another Riccati equation, which sometimes is simpler. Examples are given in
Refs. \cite{twsix}, \cite{twseven}, \cite{tweight}.

We shall go along a different way. When $B$ satisfies a Riccati equation the
same is true for $\frac aB+b$, where $a$ and $b$ are arbitrary functions of $%
t$ and $r$. Let us take first $P\equiv R^{\prime }/B$. Eq (27) becomes a
Riccati one for $P$: 
\begin{equation}
\dot P=\left( \frac A{2R}+\frac{A^{\prime }}{R^{\prime }}\right) P^2+\frac{%
A^{\prime }\dot R}{AR^{\prime }}P-\frac R{2A}\left( \frac{2\ddot R}R+\frac{%
\dot R^2}{R^2}-\frac{2\dot A\dot R}{AR}+\frac{A^2}{R^2}\right) .
\label{tweight}
\end{equation}
Next we can write 
\begin{equation}
P=H-\frac{\dot R}A.  \label{twnine}
\end{equation}
Replacing this expression in the previous equation yields a Riccati equation
for the horizon function $H$ on the star surface. It seems to be more
complicated than Eq (28). However, the opposite is true. Many of the terms
containing time derivatives of $R$ cancel each other. The final result is
rather simple: 
\begin{equation}
\dot H=\left( \frac A{2R}+\frac{A^{\prime }}{R^{\prime }}\right) H^2-\left(
\frac AR+\frac{A^{\prime }}{R^{\prime }}\right) \frac{\dot R}AH-\frac A{2R}.
\label{thirty}
\end{equation}
Only one term with $\dot R$ survives. While $A$ is related to the
four-acceleration and $R$ is the physical radius of the star as seen from
the outside, $B$ is just a metric component without direct physical meaning.
On the contrary, as we have explained, $H$ has a lot of physical
applications and it's important to have a simple equation for it. Eq (17)
shows that the redshift also satisfies a Riccati equation. However, it is
more complicated and inconvenient to work with than Eq (30).

Once again we can obtain solutions by choosing $A\left( r,t\right) $ and $%
R\left( r,t\right) $ explicitly, taking their derivatives and setting $%
r=r_\Sigma $ to obtain the coefficients of Eq (30) in explicit form. We can
also reduce it to a linear equation or a special Riccati equation without a
linear term. The Bernoulli equation does not appear here because $A\neq 0$.

Multiplying Eq (30) by $R$ one obtains a similar equation for $D\equiv RH$%
\begin{equation}
\dot D=\left( \frac A{2R}+\frac{A^{\prime }}{R^{\prime }}\right) \frac{D^2}R-%
\frac{A^{\prime }\dot R}{AR^{\prime }}D-\frac A2.  \label{thone}
\end{equation}
The expression for $B\left( t,r\right) $ follows from the definition of $H$,
Eq (10) 
\begin{equation}
B=\frac{R^{\prime }}{H-\dot R/A}.  \label{thtwo}
\end{equation}
During the collapse the radius of the star shrinks, hence, $\dot R<0$. Since 
$B$ is positive, we need $R^{\prime }>0$. This is exactly the condition for
the absence of shell crossing singularities \cite{foone}. Using Eq (30) we
can find another expression for the mass of the star. Let us multiply Eq
(12) by $A/R+A^{\prime }/R^{\prime }$, provided the latter combination does
not vanish. Utilising Eq (30) as an expression for $\dot R$, we get 
\begin{equation}
\left( \frac AR+\frac{A^{\prime }}{R^{\prime }}\right) \frac{2m}R=\frac{%
A^{\prime }}{R^{\prime }}\left( 1+H^2\right) -2\dot H,  \label{ththree}
\end{equation}
trading $\dot R$ for $\dot H$. Sometimes this formula is useful.

In the formalism described above we find collapse solutions along the
following chain. First we take positive functions $A\left( t,r\right) $ and $%
R\left( t,r\right) $, satisfying $\dot R<0$ and $R^{\prime }>0$. Then a
positive solution for $H$ is found from Eq (30) which must be a decreasing
function, starting from some $H_i$. This allows to obtain immediately the
redshift from Eq (17). Next we find the mass $m$ from Eq (12) and $B$ from
Eq (32). Now it is possible to measure the heat flow $q$ from Eq (23) and
the energy density $\mu $. Then the two luminosities and the temperature on
the star surface are given by Eqs (18,19,20). The two pressures are found
from the Einstein equations (5,6). Some of these quantities are defined on
the surface, some throughout the star. For realistic solutions we must have $%
A,B,R,R^{\prime },m,m^{\prime },q,H,\mu ,p_r>0$ and $\dot R,\dot H,\dot m<0$%
. The energy conditions should hold, too.

\section{All geodesic solutions}

The power of Eq (30) for the horizon function is best revealed in the case
of geodesic solutions. The particles of the fluid move along geodesics and
the four-acceleration $a_1$ is zero. Eq (9) shows that $A$ depends only on
time. It can be set to $1$ by a time transformation of the line element.
Then all terms with $A^{\prime }$ disappear. Eq (30) becomes 
\begin{equation}
2R\dot H=H^2-2\dot RH-1,  \label{thfour}
\end{equation}
Eq (34) has already been found \cite{twenty} from the equation for the Z
function \cite{eighteen}. Eq (31) transforms into 
\begin{equation}
\dot D=\frac 1{2R^2}D^2-\frac 12.  \label{thfive}
\end{equation}
This is a Riccati equation for $D$, but an algebraic one for $R$ and $D$ may
be taken as the generating function of the solutions. Then 
\begin{equation}
R=\frac D{\sqrt{2\dot D+1}},  \label{thsix}
\end{equation}
Now, since this equation holds on the star's surface, we can promote the
constants in it into functions of $r$ and add to the r.h.s. the term $%
G\left( t,r\right) =g\left( r\right) F\left( t,r\right) $ where $g$ and $F$
are arbitrary, as long as $R$ is positive. In addition $g\left( r_\Sigma
\right) =0$ and the term $G$ does not show up on the surface. This is a
second generating function allowed by Eq (36). We set it to zero for
simplicity. Then Eq (36) and the definition of $D$ yield 
\begin{equation}
H=\sqrt{2\dot D+1}.  \label{thseven}
\end{equation}
Eq (32) with $A=1$ gives an expression for $B$. Eq (33) becomes 
\begin{equation}
m=-R^2\dot H.  \label{theight}
\end{equation}
These relations have been found in a different way too \cite{twenty} . Thus
a zero of $H$ signals the formation of a black hole, while a zero in its
time derivative signals the burning away of the mass due to radiation and
the appearance of flat spacetime - another end of the collapse. The decrease
in the mass is given by Eq (23) with $A=1$. One can give expressions for all
stellar characteristics in terms of $D$ and its time derivatives.

This, however, is not the only generation function. One can integrate Eq
(37) and obtain an analogue of Eq (36) 
\begin{equation}
R=\frac 1{2H}\left( \int H^2dt-t\right) .  \label{thnine}
\end{equation}
Now the horizon function plays also the role of a generating function.
Further, we can replace $H$ with the redshift from Eq (17) to obtain 
\begin{equation}
R=2\left( 1+z\right) \left( \int \frac{dt}{\left( 1+z\right) ^2}-t\right) .
\label{forty}
\end{equation}
Thus the role of a third generating function is played by the redshift,
which is a physical observable.

An important issue is to restore from this general formalism the particular
solutions, found in the past \cite{twenty}. The generalized traveling waves
and self-similar solutions \cite{nineteen} may be found by choosing $D$ to
depend on an special intermediate function of $t$ and $r$. The linear in
time solution of Ref. \cite{eighteen} seems unphysical, because it cannot
have both the mass and the heat flow positive. The analogy with the
shearless case seems to break here. The other solution given in this
reference is physical, depends on several functions of $r$ and can evolve
either to a black hole or to flat spacetime, depending on how we choose
these functions.

\section{Special Riccati equation}

Let $A$ and $R$ satisfy on the star's surface $\Sigma $ the equation 
\begin{equation}
\frac AR+\frac{A^{\prime }}{R^{\prime }}=0.  \label{foone}
\end{equation}
This eliminates the linear term in Eq (30) and it becomes 
\begin{equation}
\dot H=\left( \frac A{2R}+\frac{A^{\prime }}{R^{\prime }}\right) H^2-\frac
A{2R}.  \label{fotwo}
\end{equation}
Eq (41) provides a relation between $A$ and $R$: 
\begin{equation}
R=\frac{K\left( t\right) }A,  \label{fothree}
\end{equation}
where $K\left( t\right) $ is an arbitrary positive function. It makes Eq
(42) integrable 
\begin{equation}
\frac{dH}{1+H^2}=-\frac{A^2}{2K}dt,  \label{fofour}
\end{equation}
which yields the solution 
\begin{equation}
H=tg\frac 12\int_t^{t_f}\frac{A^2}Kdt.  \label{fofive}
\end{equation}
Here $t_f$ is the end time of collapse where $H=0$ and a black hole, covered
by a horizon is formed. Eq (44) gives 
\begin{equation}
\frac{2R}A=-\frac{1+H^2}{\dot H}.  \label{fosix}
\end{equation}
We may take $A$ and $H$ as independent arbitrary functions and find $R$ from
this equation. Let $A=A\left( r\right) $. Then $R$ is a solution with
separated variables and the above equation yields upon differentiation 
\begin{equation}
\frac{2\dot R}A=-2H+\left( 1+H^2\right) \frac{\ddot H}{\dot H^2}.
\label{foseven}
\end{equation}
Replacing this expression in Eq (12) for the mass we obtain 
\begin{equation}
\frac{2m}R=1-3H^2+\left( 1+H^2\right) \frac{H\ddot H}{\dot H^2}.
\label{foeight}
\end{equation}
Thus the compactness parameter $u=m/R$ depends only on $H$ and its time
derivatives. Now, since on the star's surface $A\left( r_\Sigma \right) $
and $A^{\prime }\left( r_\Sigma \right) $ are some constants, $H_\Sigma $
becomes the only generating function for this class of solutions. One must
be careful to choose it in such a way that all above mentioned inequalities
are satisfied in order to obtain physically realistic solutions. We shall
demonstrate that this is not so easy.

Let us choose the simplest case when $\ddot H_\Sigma =0$. Then 
\begin{equation}
H_\Sigma =c_1-c_2t,  \label{fonine}
\end{equation}
where $c_i$ are positive constants and $c_1>c_2$ . The collapse starts at
some initial time $t_i$ and ends at $t_f=c_1/c_2$. $H_\Sigma $ is positive
and decreasing, $\dot H_\Sigma $ is negative. Hence $R_\Sigma $ in Eq (46)
is positive and can be written as 
\begin{equation}
R_\Sigma =\frac{A_\Sigma }{2c_2}\left( 1+H_\Sigma ^2\right) .  \label{fifty}
\end{equation}
It decreases with time, therefore $\dot R_\Sigma <0$. Eqs (48,50) give for
the mass 
\begin{equation}
m_\Sigma =\frac{A_\Sigma }{4c_2}\left( 1-2H_\Sigma ^2-3H_\Sigma ^4\right) .
\label{fione}
\end{equation}
This is positive as long as $H_\Sigma <1/\sqrt{3}$. This may be arranged by
choosing $c_i$ appropriately for any $t_i$. However, the time derivative of
the mass is 
\begin{equation}
\dot m_\Sigma =A_\Sigma H_\Sigma \left( 1+3H_\Sigma ^2\right) ,
\label{fitwo}
\end{equation}
which is always positive. Thus $q_\Sigma <0$ and energy is pumped into the
star from the outside until it turns into a black hole when $2u_\Sigma =1$.
This does not seem realistic. The study of concrete solutions will be
continued elsewhere.

\section{Linear equation}

The quadratic term in Eq (30) disappears and it becomes a linear equation
when 
\begin{equation}
\frac A{2R}+\frac{A^{\prime }}{R^{\prime }}=0.  \label{fithree}
\end{equation}
It leads to the relation 
\begin{equation}
R=\frac{K\left( t\right) }{A^2},  \label{fifour}
\end{equation}
where, as before, $K\left( t\right) $ is an arbitrary positive function.
This case was discussed in Ref. \cite{twsix} and for $B$ it leads to a
Bernoulli equation with very complicated solution. Here Eq (30) simplifies
considerably 
\begin{equation}
2R\dot H+\dot RH+A=0  \label{fifive}
\end{equation}
and is linear both in $H$ and in $R$. Let us suppose for simplicity that
collapse starts at $t=0$ when a non-trivial heat flow $q$ appears, due to
processes taking place in the star, being static before that time. The
solution of the general linear differential equation 
\begin{equation}
g\dot H=f_1H+f_0  \label{fisix}
\end{equation}
can be written in the form 
\begin{equation}
H=e^F\left( H_i+\int_0^te^{-F}\frac{f_0}gdt\right) ,  \label{fiseven}
\end{equation}
where 
\begin{equation}
F=\int_0^t\frac{f_1}gdt,  \label{fieight}
\end{equation}
clearly showing that the initial value of the horizon function is $H_i$. It
differs slightly from the formula usually given in handbooks \cite{fotwo}.
For Eq (55) we get 
\begin{equation}
F=-\frac 12\ln \frac R{R_i},\quad R_i=R\left( t=0\right) ,  \label{finine}
\end{equation}
\begin{equation}
H=\sqrt{\frac{R_i}R}\left( H_i-\frac 1{2\sqrt{R_i}}\int_0^t\frac{A^2}{\sqrt{K%
}}dt\right) .  \label{sixty}
\end{equation}
Provided $A$ and $K$ are given, we can find $R,H,B,m,q$ and the other
characteristics of the star.

In the shearless case a simple linear in time solution for $R=rB$ leads to
''eternal collapse'' without horizon \cite{seven}, \cite{eleven}, \cite
{twelve}. It is natural to study similar solution in the case with shear,
that we are discussing. Thus, let us take 
\begin{equation}
R_\Sigma =R_i-bt  \label{sione}
\end{equation}
on the star's surface, with positive constants $R_i,b$. Let $A=A\left(
r\right) $. Then the integration in Eq (60) yields 
\begin{equation}
H_\Sigma =\frac{A_\Sigma }b-\sqrt{\frac{R_i}{R_\Sigma }}\left( \frac{%
A_\Sigma }b-H_i\right) =c-\sqrt{\frac{R_i}{R_\Sigma }}\varepsilon ,
\label{sitwo}
\end{equation}
where $A_\Sigma /b\equiv c$, and $c-H_i\equiv \varepsilon $. Here $c$ and $%
\varepsilon $ are positive constants. Obviously $\dot R_\Sigma <0$ and $\dot
H_\Sigma <0$. At some $t_f$ we get $H_{f\Sigma }=0$ for the corresponding $%
R_{f\Sigma }$, that is, a black hole appears: 
\begin{equation}
R_{f\Sigma }=\frac{\varepsilon ^2}{c^2}R_i  \label{sithree}
\end{equation}

Eq (12) gives for the mass 
\begin{equation}
2m_\Sigma =-R_i\varepsilon ^2+2\sqrt{R_iR_\Sigma }\varepsilon \frac{1+c^2}%
c-\left( 1+c^2\right) R_\Sigma  \label{sifour}
\end{equation}
and its time derivative 
\begin{equation}
2\dot m_\Sigma =\left( 1+c^2\right) \dot R_\Sigma \left( \frac \varepsilon c%
\sqrt{\frac{R_i}{R_\Sigma }}-1\right) .  \label{sifive}
\end{equation}
The second bracket is negative because of Eq (63) and $R_\Sigma >R_{f\Sigma
} $. Hence, the mass of the star is a monotonously increasing function, like
in the previous section.

Now we can prove that at least $m_\Sigma $ can be chosen positive. First, at 
$R_{f\Sigma }$ we get 
\begin{equation}
2m_{f\Sigma }=\frac{\varepsilon ^2R_i}{c^2},  \label{sisix}
\end{equation}
which is positive. Second, at the beginning of collapse we have $R_\Sigma
=R_i$ and 
\begin{equation}
2m_{\Sigma i}=R_i\left[ 2\varepsilon \frac{1+c^2}c-\varepsilon ^2-\left(
1+c^2\right) \right] .  \label{siseven}
\end{equation}
This expression is positive when 
\begin{equation}
H_i^2+\frac 2cH_i-1<0.  \label{sieight}
\end{equation}
One possible solution of this inequality is 
\begin{equation}
H_i=\frac c4,\quad c<2\sqrt{2}.  \label{sinine}
\end{equation}

We conclude that the simplest solution in this section is unphysical because 
$\dot m_\Sigma >0$.

\section{Discussion}

The main purpose of this paper was to reformulate the main junction equation
in terms of the physically important function $H$. This is possible because,
originally, this condition represents a Riccati equation for the metric
component $B$. It is known that under a fractional linear transformation of
the unknown function the Riccati equation for it preserves its nature, but
the coefficients attached to the different terms change \cite{fotwo}. Since $%
H$ is such a transformation of $B$, it also satisfies a Riccati equation.
One expects this to be useless, because the coefficients seem to become much
more complicated. Due to reasons, unknown to us, there are miraculous
cancellations and most of the terms cancel each other. The resulting
equation is simpler and directly guides the behaviour of $H$ and its zero,
where a horizon and a black hole appear.

Eq (17) gives a simple relation between the observable redshift and the
horizon function $H$, which holds in any case. It represents a fractional
linear transformation, this time of $H$. Hence, the redshift also satisfies
a Riccati equation. More generally, every statement about $H$ can be
translated into a statement about the redshift. Eq (19) shows that the
square of $H$ transforms the surface luminosity (depending mainly on the
heat flow) into the luminosity at infinity and serves as a measure of its
faintness. The horizon function also serves as a main ingredient of the mass
formula and the compactness parameter. The time derivative of the mass gives
a simpler expression for the heat flow, which is fundamental for the surface
temperature and luminosity of the star.

In this different approach we again start with analytic expressions for $A$
and the luminous radius of the star $R$ and then solve analytically the
equation for $H$. A signal for its great potential is that for geodesic
fluid motion all solutions may be found explicitly with the help of two
generating functions. One of them can be chosen to be $D\equiv RH$, $H$ or
the redshift $z$ \cite{twenty}. Thus, observing the time evolution of the
redshift, e.g., is enough to restore the evolution of the star
characteristics on the surface.

In the general case, setting certain coefficient to zero changes the Riccati
equation into an integrable equation. This requires a simple relation
between $A$ and $R.$ In some cases $H$ plays alone the role of a generating
function, like in the geodesic case, (see Eq (46) with $A=A\left( r\right) $%
). The simplest solutions, unfortunately, are astrophysically unrealistic
(the heat flow is negative), but still help to see how a black hole appears.
We relegate the presentation of realistic solutions to future work.

This different approach works when $A,B$ and $R$ are independent apriori. In
the shearless case $R=rB$ and the junction equation is not Riccati, but a
more complex one. Usually the absence of shear is accepted in order to
simplify the problem. We see that the opposite is true. The general case
with shear is simpler than the shearless one.

\end{document}